\documentclass[aps,pra,showpacs,superscriptaddress,groupedaddress]{revtex4}  

\usepackage{graphicx}   
\usepackage{bm}             
\usepackage{amssymb}   
\usepackage{epsfig}
\usepackage{color}

\begin{document}


\title{Vibrationally resolved NO dissociative excitation cross sections by electron impact}

\author{V.~Laporta}
\email{vincenzo.laporta@istp.cnr.it}
\affiliation{Istituto per la Scienza e Tecnologia dei Plasmi, CNR, Bari, Italy}

\author{J.~Tennyson}
\affiliation{Department of Physics and Astronomy, University College London, London, UK}

\author{I.F.~Schneider}
\affiliation{Laboratoire Ondes et Milieux Complexes, CNRS--Universit{\'{e}} Le Havre Normandie, Le Havre, France}
\affiliation{Laboratoire Aim{\'{e}} Cotton, CNRS--Universit{\'{e}} Paris-Saclay, ENS Cachan, Orsay, France}

\begin{abstract}
A theoretical investigation of the dissociative excitation by electron impact on the NO molecule is presented, aiming to make up for the lack of data for this process in the literature. A full set of vibrationally-resolved cross sections and corresponding rate coefficients are calculated using  the Local-Complex-Potential approach and five resonant states of NO$^-$. 
\end{abstract}

\pacs{xxxxx}

\maketitle


Nitric oxide (NO) molecule is one of the minor components of terrestrial 
atmosphere. Generated in atmospheric plasma from chemical reactions of nitrogen 
with oxygen, NO and its radicals are very important in many industrial 
technologies~\cite{doi:10.1080/0144235X.2016.1179002, 0963-0252-22-2-025008, 
Campbell_2012, Motapon-PSST-2006, doi:10.1029/2002JA009458} and play a
key role in the combustion  of  fossil 
fuels~\cite{SLAVCHOV2020117218, doi:10.1080/00102206908952211}. Of the 
nitrogen oxide compounds, the so-called $\mathrm{N}\mathrm{O}x$ gasses, the NO molecule has the 
greater impact on environment and on pollution caused by human 
activities~\cite{Kreuzer45, Spicer:1977aa}.  

In order to make kinetic plasma models involving nitric oxide, many sets 
of molecular data~\cite{vivie:3028, LOCHT1970379, C0CP01067G}, spectroscopic 
properties~\cite{10.1093/mnras/stx1211, 0022-3700-10-4-006, HARGREAVES201935} and reaction rate 
coefficients~\citep{doi:10.1063/1.5114722, doi:10.1063/1.4961372, ZECCA2003205, 
Brunger2002215} are available in the literature  but, in 
spite of its importance, none of them provide complete data on electron impact dissociation. More specifically, 
rate constants for electron-NO reactions, both 
theoretical ~\cite{0022-3700-10-4-006,0963-0252-21-5-055018, PhysRevA.71.052714, PhysRevA.69.062711} and 
experimental~\cite{0953-4075_38_5_011, PhysRevLett.90.203201, 
Josic2001318, Krishnakumar_1988, PhysRevA.10.1633}, exist only for vibrational 
excitation and dissociative electron attachment processes at low-energy. A recent compilation
of all the known electron collision cross sections is given by Song {\it et al.}~\citep{doi:10.1063/1.5114722}.

To fill this gap,  we present calculations -- based on the 
formalism used previously~\cite{0963-0252-21-5-055018} -- of vibrational state 
resolved cross sections and the corresponding rate coefficients for 
dissociative excitation (DE) of NO by electron impact, \textit{i.e.}:
\begin{equation}
e + \mathrm{NO}(\mathrm{X}\,^2\Pi; v) \to \mathrm{NO}^- \to e + \mathrm{NO}(\mathrm{X}\,^2\Pi; \epsilon_c) \to e 
+\mathrm{N}(^4\mathrm{S}) + \mathrm{O}(^3\mathrm{P})\,.\label{eq:DEprocess}
\end{equation}
We consider electron collision energies where 
numerous NO$^-$ resonances exist, direct dissociation is negligible and the DE reaction is dominated by resonant
processes~\cite{0034-4885-31-2-302, Domcke199197}. 
Our aim is to cover a large range of incident electron energies, so we take into 
account five resonance states of NO$^-$:  the three low-lying states of  
$^3\Sigma^-$, $^1\Sigma^+$ and $^1\Delta$ symmetries and two higher ones, with 
$^3\Pi$ and $^1\Pi$ symmetry, which lie close to the NO dissociation threshold. In the 
following, we number these resonances by $r=1,\ldots,5$, respectively. The vibrational excited states of which converge  on
the $\mathrm{N}(^4\mathrm{S}) + \mathrm{O}(^3\mathrm{P})$
dissociation limit of Eq.~(\ref{eq:DEprocess}), due to their symmetry,   have small oscillator strengths and short lifetimes~\cite{vivie:3028, Brzozowski_1974, Camilloni1987} so their influence on the DE process  can be neglected.


NO is a stable, open-shell molecule with a $^2\Pi$ ground electronic state, and 
a very accurate theoretical treatment of the electron-NO scattering therefore 
needs to take into account the spin-dependence of the process. However, the 
spin-orbit coupling effects are only important at very low energies~\cite{0953-4075_38_5_011}
and, consequently, in the following we will neglect them.

We start by briefly describing the theoretical model used to calculate the cross 
sections for the process (\ref{eq:DEprocess}), restricting ourselves to the 
major equations of the Local-Complex-Potential (LCP) model.  For a 
comprehensive treatment of the resonant collisions, we refer to the seminal 
articles~\cite{0034-4885-31-2-302, Domcke199197, PhysRevA.20.194}. The LCP 
approach was used to calculate the low-energy vibrational 
excitation of NO~\cite{0963-0252-21-5-055018} and CO~\cite{0963-0252-21-4-045005} by electron impact and the DE of 
oxygen molecule~\cite{0963-0252-22-2-025001, PhysRevA.91.012701}, which  gave results in
good agreement with experiment.

In the LCP model, the DE cross section for an NO molecule initially in 
vibrational level $v$ by an incident electron of energy $\epsilon$ is given 
by~\cite{PhysRevA.20.194}:
\begin{equation}
\sigma_{v}(\epsilon) = \sum_{r=1}^5 \frac{2S_r+1}{(2S+1)\,2} \frac{g_r}{g\,2} 
\frac{64\,\pi^5\,m^2}{\hslash^4} 
\int_{\epsilon^{th}_v}^{\epsilon^{max}_v}d\epsilon_c \frac{k'}{k}\left|\langle 
\chi_c|\mathcal{V}_r|\xi^r_v \rangle\right|^2\,, \label{eq:DExsec}
\end{equation}
where $2S_r+1$ and $2S+1$ are the spin-multiplicities of the resonant  
anion state and of the neutral target state 
respectively, $g_r$ and $g$ represent the corresponding degeneracy factors, 
$k$($k'$) are the incoming (outgoing) electron momenta, $m$ is the electron 
mass, $\chi_c$ stands for the continuum wave function of NO asymptotically 
converging to $\mathrm{N}(^4\mathrm{S}) + \mathrm{O}(^3\mathrm{P})$, and 
$\xi^r_v$ is the resonance nuclear-motion wave function which is obtained for each resonace as the solution of 
Schr\"dinger equation:
\begin{equation}
\left[ -\frac{\hslash^2}{2\mu}\frac{d^2}{dR^2} + V_r^-(R)  - 
\frac{i}{2}\Gamma_r(R) - E \right]\xi^r_v(R) = -\mathcal{V}_r(R)\,\chi_v(R)\,, 
\hspace{1cm}r=1,\ldots,5 \,. \label{eq:res_wf}
\end{equation}
where $\mu$ is the NO reduced mass, $V_r^-$ and 
$\Gamma_r$ represent the potential energies and the autoionization widths 
respectively for the five resonant NO$^-$ states included in the calculation, 
$\chi_v$ is the wave function of the initial vibrational state of NO with energy 
$\epsilon_v$, and $E=\epsilon + \epsilon_v$ is the total energy of the system. 
Following Ref.~\cite{PhysRevA.20.194}, the widths $\Gamma_r$ were 
expressed as:
\begin{equation}
\Gamma_r = c_r \left( V_r^- - V_0 \right)^{l_r+\frac12}\,H\left(V_r^- - 
V_0\right)\,, \hspace{1cm}r=1,\ldots,5 \,, \label{eq:gamma}
\end{equation}
where $V_0$ is the NO potential energy, $H$ is a Heaviside step-function and 
$l_r$ is the angular momentum of the lowest contributing partial wave associated 
with the incident electron. Experimental results show that the dominant contribution come from the $p$-wave~\cite{PhysRevA.69.062711} so $l_r$ was set to one. In order to reproduce the 
positions and widths of the peaks in the low-energy region of the experimental 
cross sections, the constants $c_r$ in Eq.~(\ref{eq:gamma}) were introduced as 
phenomenological external parameters as reported in the 
paper~\cite{0963-0252-21-5-055018}. Their values are given in 
Table~\ref{tab:params}.
In Eqs.~(\ref{eq:DExsec}) and (\ref{eq:res_wf}), $\mathcal{V}_r$ is the 
discrete-to-continuum coupling given by~\cite{PhysRevA.71.052714}:
\begin{equation}
\mathcal{V}_r^2 = 
f_r^{2l_r+1}\,\frac{\hslash}{2\pi}\frac{\Gamma_r}{\sqrt{2\,m\left(V_r^- - 
V_0\right)}}\,, \hspace{1cm}r=1,\ldots,5 \,, \label{eq:couplingVr}
\end{equation}
where $f_r$ is the so-called penetration factor introduced to reproduce the 
correct dependence of the cross sections on the energy near the excitation 
threshold, namely,
\begin{equation}
f_r^2(\epsilon) = \left\{
\begin{array}{ll}
\epsilon/\left(V_r^- -V_0\right) & \textrm{if~} \epsilon < (V_r^- -V_0)
\\
1 & \textrm{otherwise}
\end{array}\right. \,, \hspace{1cm}r=1,\ldots,5 \,.
\end{equation}
The sum in the cross section formula~(\ref{eq:DExsec}) runs over all the five 
resonance states of the  NO$^-$ anion involved in the 
dissociation process. The integral extends to the continuum part of the NO 
potential from the dissociation threshold energy $\epsilon_v^{th}$ corresponding 
to the vibrational level $v$ up to $\epsilon_v^{max} = \epsilon_v^{th} + 10$~eV. 
The spin-statistical factors are listed in Table~\ref{tab:params}.
\begin{table}
\centering
\caption{Summary on the parameters used in the theoretical model: reduced mass 
$\mu$; Morse parameters for NO and NO$^-$ potential energies; the coefficients 
$c_r$ entering in Eq.~(\ref{eq:gamma}) and the spin-statistical factors 
$\frac{2S_r+1}{(2S+1)\,2} \frac{g_r}{g\,2}$. \label{tab:params} }
\begin{tabular}{|c|c|c|c|c|c|c|}
\hline
 & ~~NO$(\textrm{X }^2\Pi)$~~ & ~~NO$^-(^3\Sigma^-)$~~ & ~~NO$^-(^1\Delta)$~~ & 
~~NO$^-(^1\Sigma^+)$~~ & ~~NO$^-(^3\Pi)$~~ & ~~NO$^-(^1\Pi)$~~ \\
\hline
$\mu$ (a.u.)      &  \multicolumn{6}{|c|}{13614.1}\\
\hline
$D_e$ (eV)         & 6.610  & 5.161    & 5.90  & 5.60  & 1.85  & 0.729 \\
$R_e$ (a.u.)        & 2.175  & 2.393   & 2.38   & 2.37  & 3.18  & 3.14 \\
$\alpha$ (a.u.)   & 1.48    & 1.20      & 1.18  & 1.20  & 0.75  & 1.049 \\
$W$ (eV)            & 0        & -0.015   & 0.775 & 1.08  & 5.70  & 6.80 \\
\hline
$c_r$ (eV$^{0.5-l_r}$)   & --    & 0.81        & 0.60        & 0.50 & 0.023 & 
0.013 \\
\hline
spin-stat & -- & $\frac{3}{8}$ & $\frac{1}{4}$ & $\frac{1}{8}$ & $\frac{3}{4}$ & 
 $\frac{1}{4}$ \\
\hline
\end{tabular}
\end{table}

\begin{figure}
\centering
\includegraphics[scale=.39]{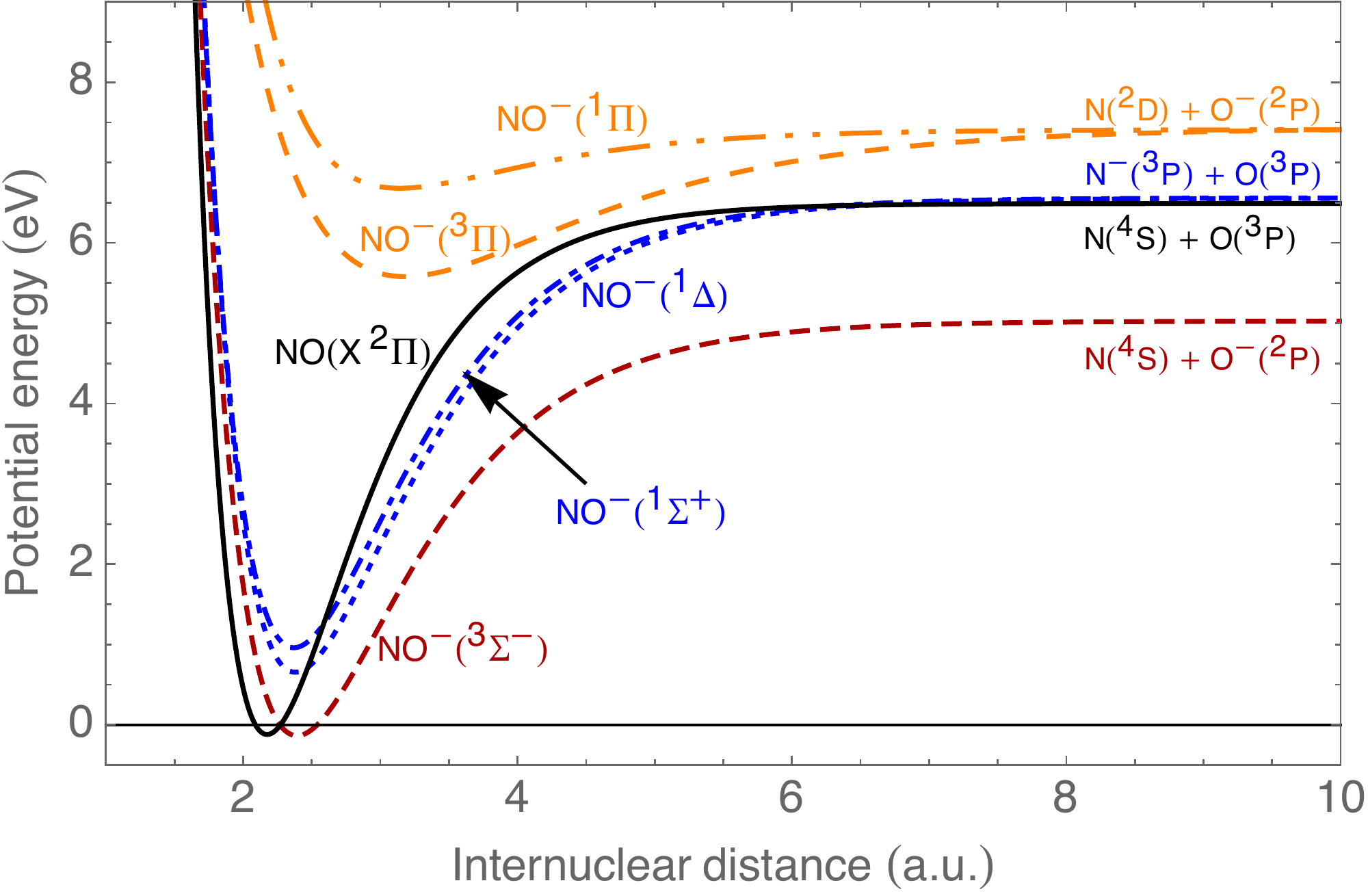} \hspace{.5cm} 
\includegraphics[scale=.4]{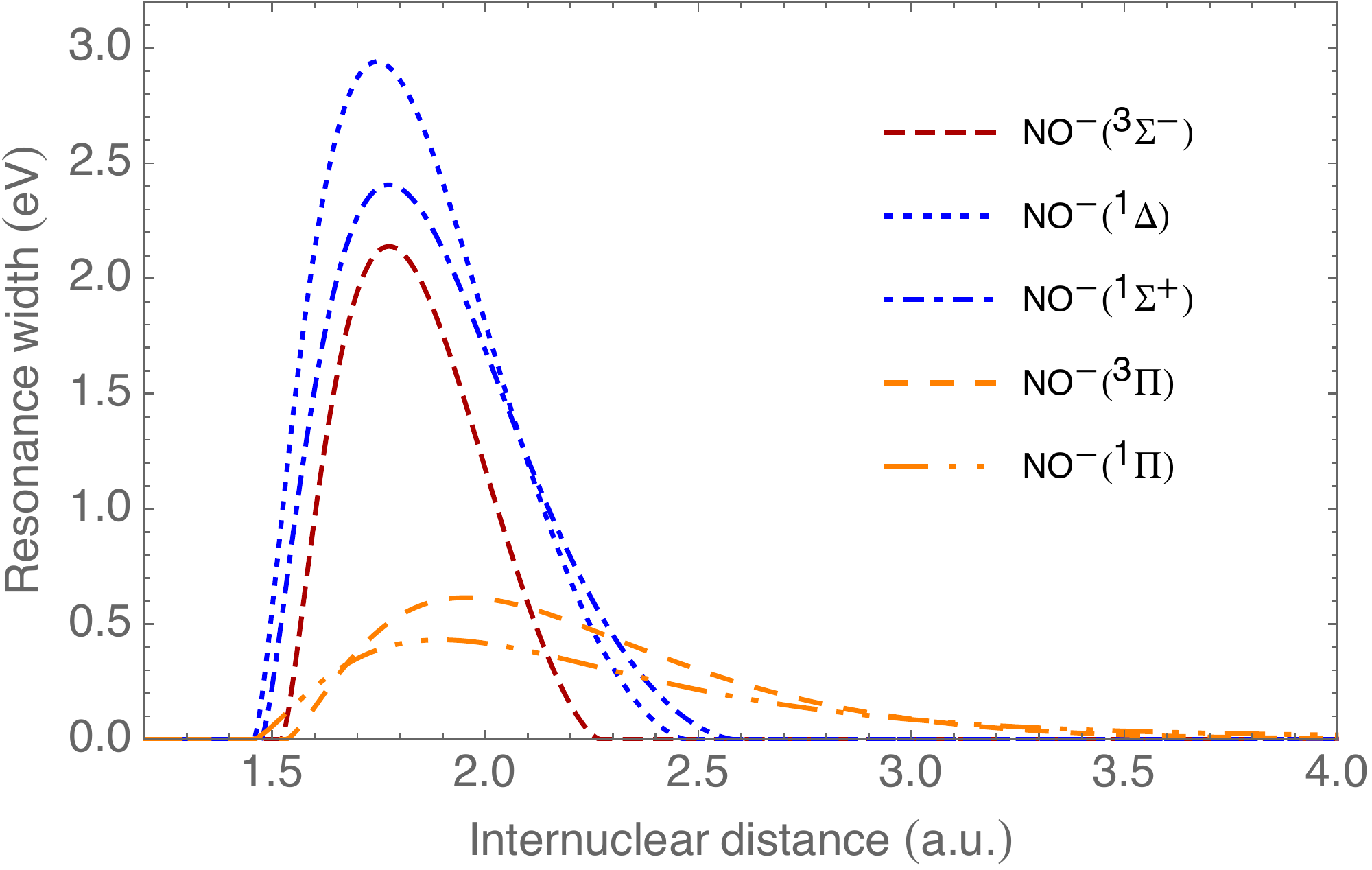} 
\caption{(Plot on the left) Potential energy curves for the ground electronic 
state of NO molecule (solid line) and for the five NO$^-$ resonances (broken 
lines) included in the calculations; (Plot on the right) The corresponding 
widths of the resonances. \label{fig:NOpot}}
\end{figure}

In the model shown above, the potentials $V_0(R)$ and $V_r^-(R)$ are
expressed as a standard Morse function $U(R) = 
D_e\left[1-e^{-\alpha(R-R_e)}\right]^2+W$ whose parameters, for the NO molecule 
and for the three low-lying resonances $^3\Sigma^-$, $^1\Sigma^+$ and 
$^1\Delta$, were determined by a fit procedure explained in 
~\cite{0963-0252-21-5-055018}. In order to take into account the recent results 
presented in~\cite{C0CP01067G}, the asymptotes for the singlet states 
$^1\Sigma^+$ and $^1\Delta$ have been shifted to the correct threshold 
$\mathrm{N}^-(^3\mathrm{P}) + \mathrm{O}(^3\mathrm{P})$. We have checked, and 
the results in ~\cite{0963-0252-21-5-055018} are not affected by these changes. 
Analogously, the $^3\Pi$ and $^1\Pi$ symmetry parameters were obtained by a fit 
to the data presented in~\cite{LOCHT1970379} and the
\textit{ab-initio} R-Matrix results in the Ref.~\cite{C0CP01067G}. All Morse 
parameters are summarized in Table~\ref{tab:params}.  The  NO ground state
potential energy curve, as 
well as those for the five NO$^-$ resonances and their corresponding autoionization 
widths are reported in Fig.~\ref{fig:NOpot}. Table~\ref{tab:NOviblev} contains 
the list of the vibrational levels supported by the NO molecule.
\begin{table}
\centering
\caption{Energies of the vibrational levels of the electronic ground state of 
the NO molecule. $D_0=6.490$~eV \label{tab:NOviblev}}
\begin{tabular}{cccccc}
\hline
~~~$v$~~~ & $~~\epsilon_{v}$~(eV)~~ & ~~~$v$~~~ & $~~\epsilon_{v}$~(eV)~~& 
~~~$v$~~~ & $~~\epsilon_{v}$~(eV)~~\\
\hline
  0  &     0.000  &   18  &     3.581  &   36  &     5.745 \\ 
  1  &     0.236  &   19  &     3.739  &   37  &     5.823 \\ 
  2  &     0.468  &   20  &     3.892  &   38  &     5.897 \\ 
  3  &     0.695  &   21  &     4.040  &   39  &     5.967 \\ 
  4  &     0.918  &   22  &     4.185  &   40  &     6.033 \\ 
  5  &     1.137  &   23  &     4.324  &   41  &     6.094 \\ 
  6  &     1.351  &   24  &     4.460  &   42  &     6.150 \\ 
  7  &     1.561  &   25  &     4.591  &   43  &     6.203 \\ 
  8  &     1.767  &   26  &     4.718  &   44  &     6.251 \\ 
  9  &     1.968  &   27  &     4.840  &   45  &     6.294 \\ 
10  &     2.164  &   28  &     4.958  &   46  &     6.333  \\ 
11  &     2.357  &   29  &     5.072  &   47  &     6.368 \\ 
12  &     2.545  &   30  &     5.181  &   48  &     6.399 \\ 
13  &     2.729  &   31  &     5.286  &   49  &     6.425 \\ 
14  &     2.908  &   32  &     5.386  &   50  &     6.446 \\ 
15  &     3.083  &   33  &     5.483  &   51  &     6.464 \\ 
16  &     3.253  &   34  &     5.574  &   52  &     6.477 \\ 
17  &     3.419  &   35  &     5.662  &   53  &     6.485 \\ 
\hline
\end{tabular}
\end{table}

\begin{figure}
\centering
\includegraphics[scale=.28]{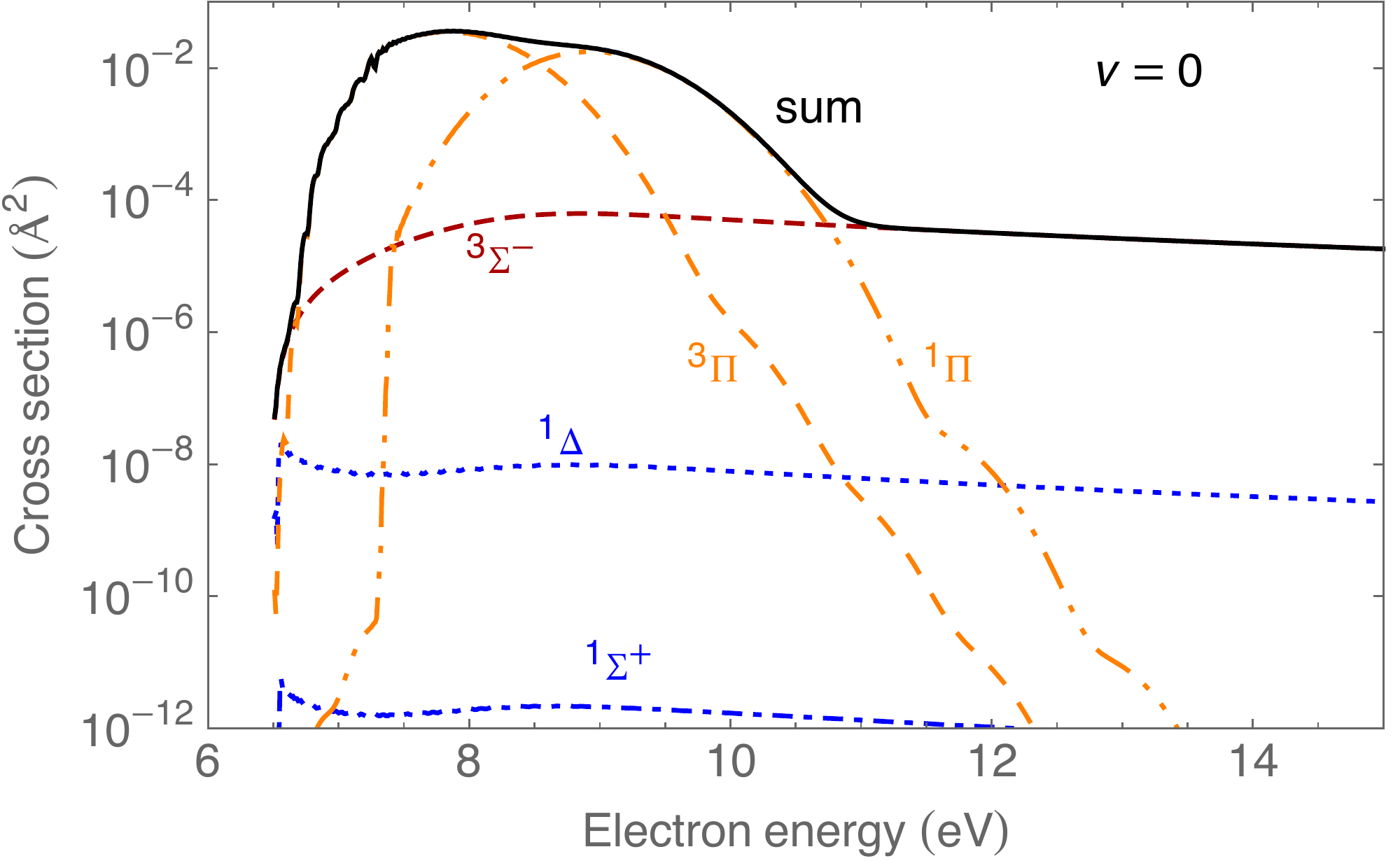} 
\includegraphics[scale=.28]{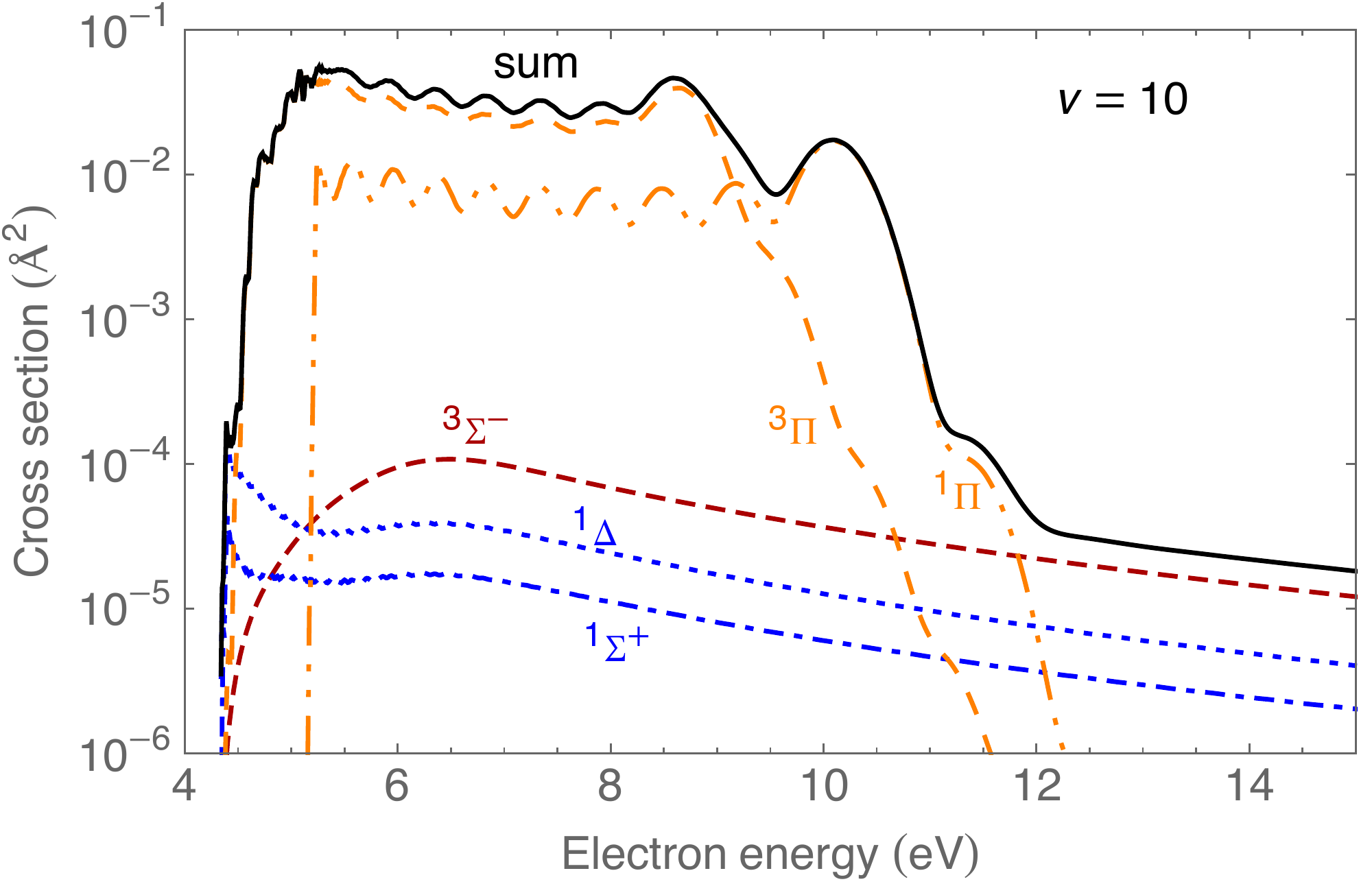} 
\includegraphics[scale=.28]{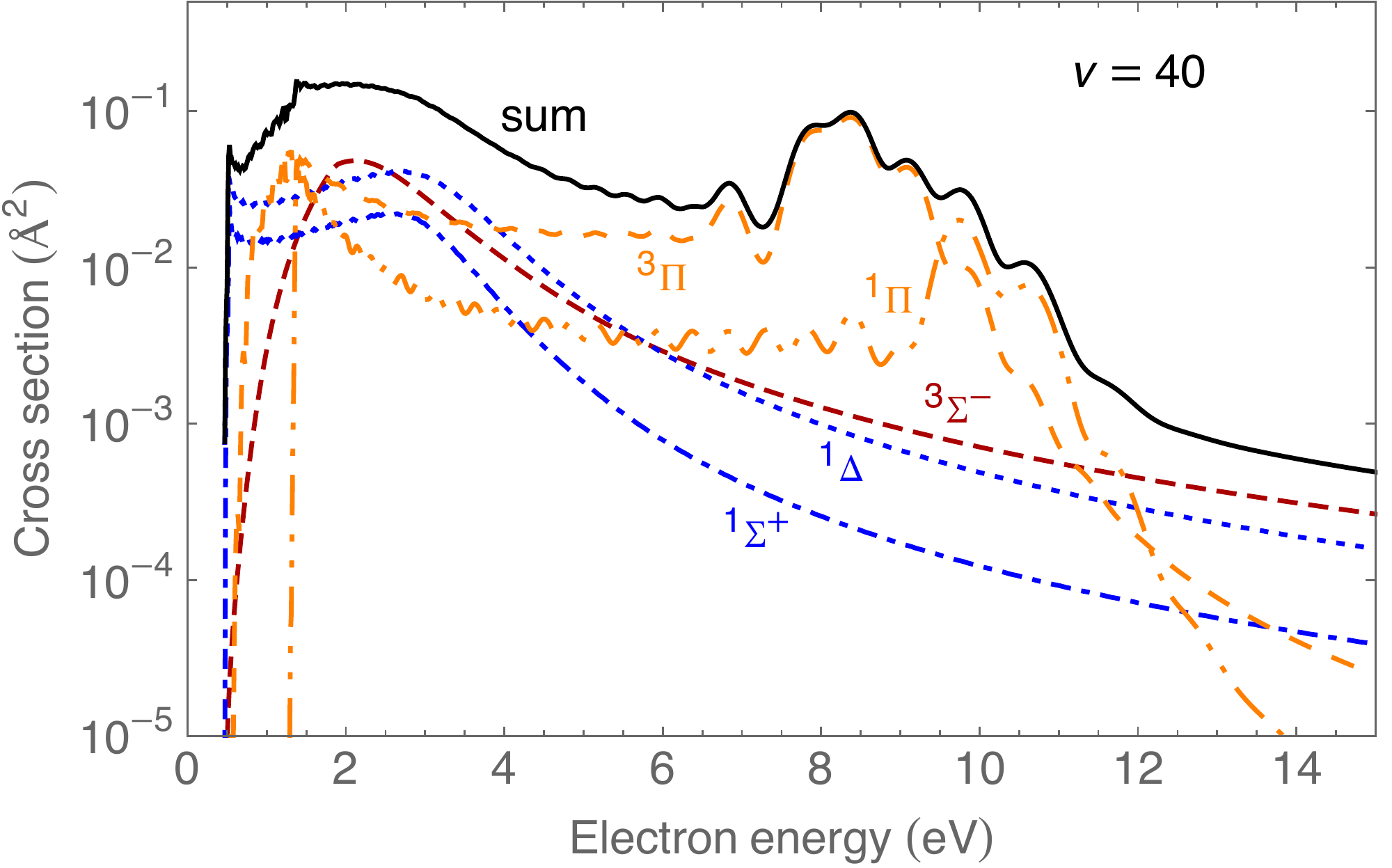} 
\caption{Contributions coming from the five resonant states (broken lines, same 
colors as in Fig.~\ref{fig:NOpot}) to the total dissociative excitation cross section (solid line) for 
$v=0$ (plot on the left), $v=10$ (plot on the middle) and $v=40$ (plot on the 
right).  \label{fig:xsecDEsym}}
\end{figure}
\begin{figure}
\centering
\includegraphics[scale=.4]{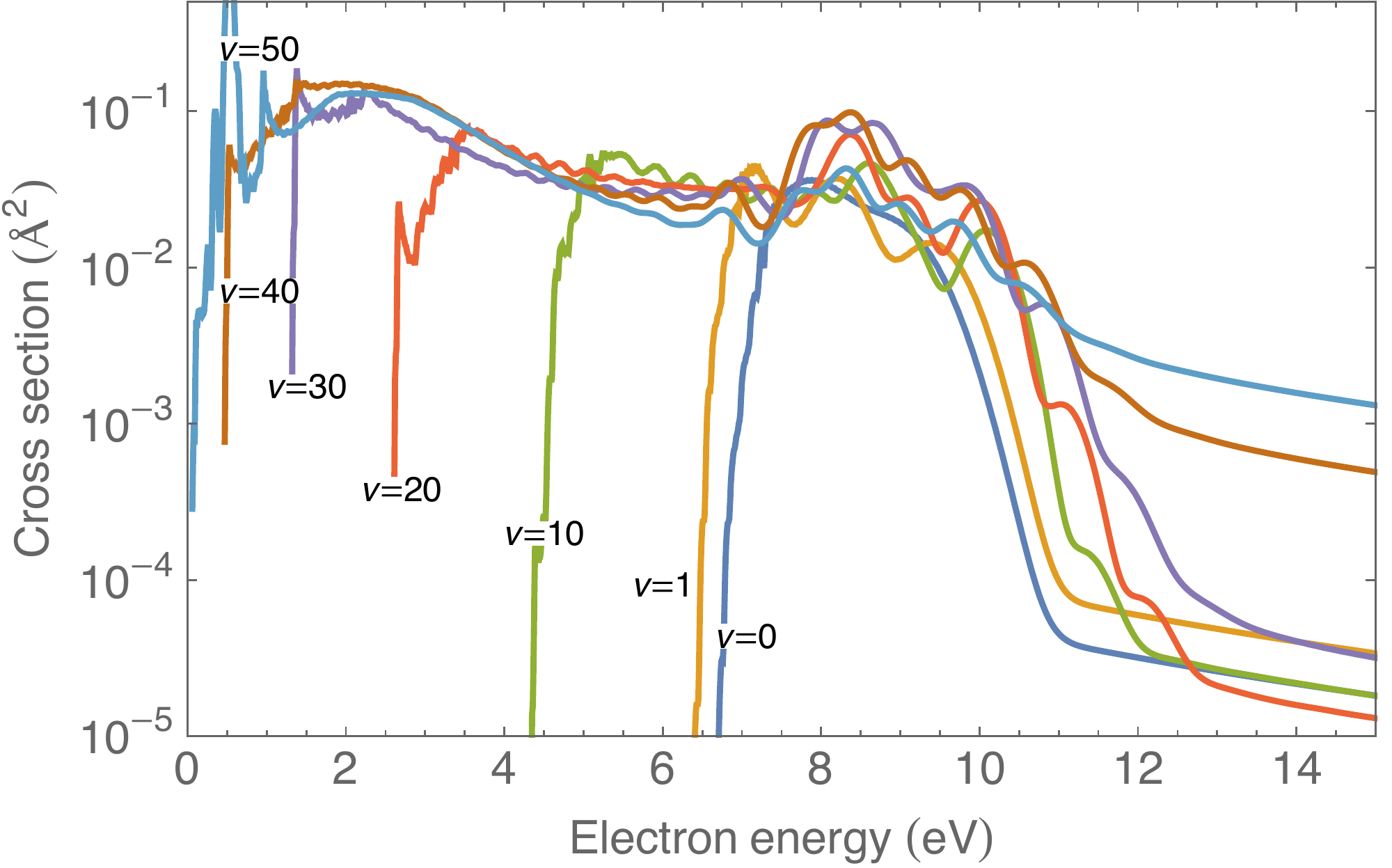} \hspace{.5cm} 
\includegraphics[scale=.4]{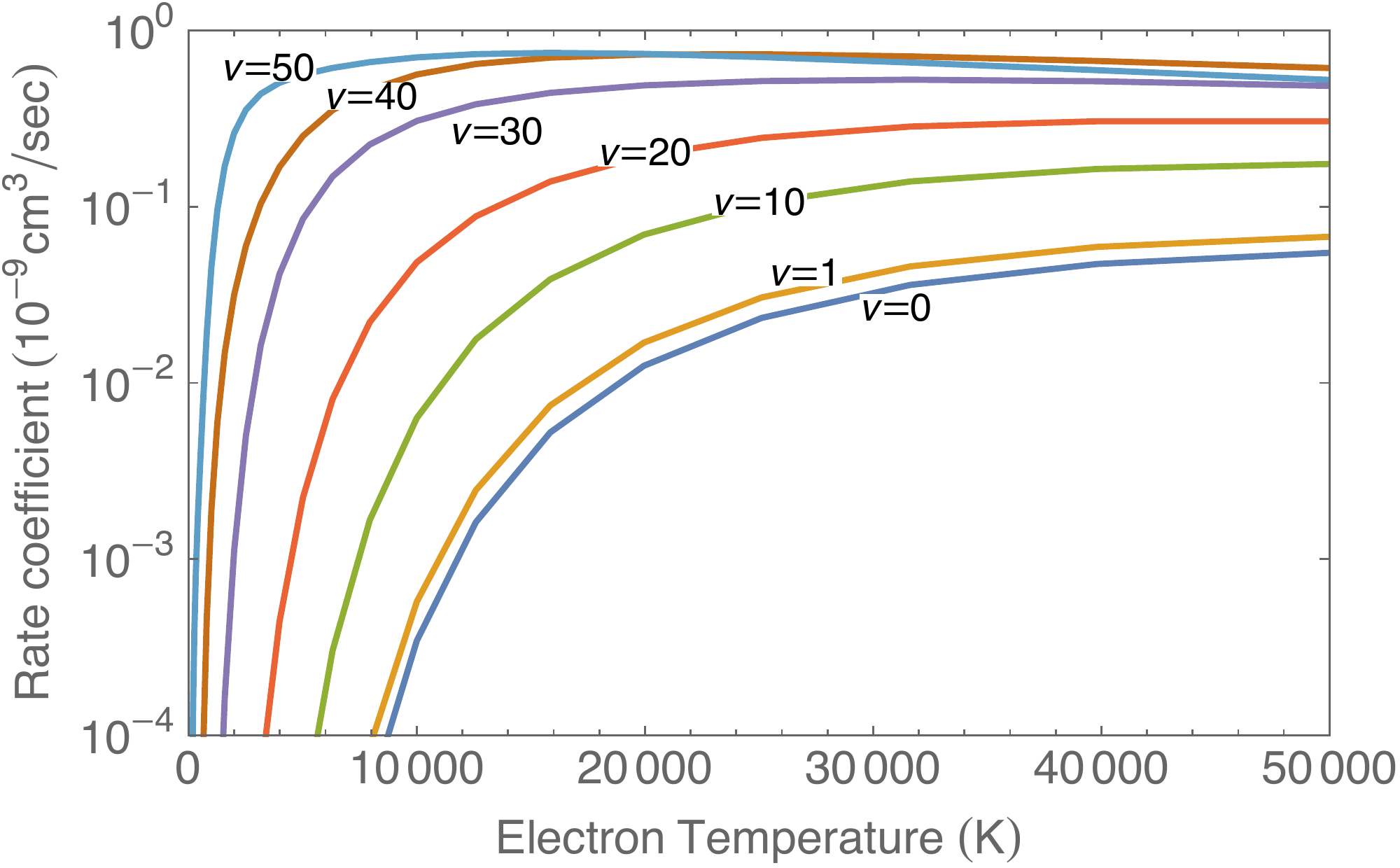} 
\caption{Vibrational state-resolved cross sections and the corresponding rate 
coefficients for NO dissociative excitation by electron impact. \label{fig:xsecDE}}
\end{figure}

Figures~\ref{fig:xsecDEsym} and \ref{fig:xsecDE} summarize the results of the present letter.
The cross sections were computed up to 15 eV, at which point  they drop off and become negligible,
yo cover   temperatures up to 50000 K for the reaction rates, relevant for the applications mentioned above.

Figure~\ref{fig:xsecDEsym} contains the results of cross sections for the 
process of dissociation in (\ref{eq:DEprocess}) for three specific vibrational 
levels of NO molecule. Partial contributions coming from the five resonances as 
well as the sum are shown. Some features can be noticed: (i) Basically, as 
expected, for all cases, the major contribution to the total cross section comes 
from the $^3\Pi$ and $^1\Pi$ resonances due to its closeness to the dissociation 
threshold, whereas the $^1\Sigma^+$ and $^1\Delta$ resonances, in general, make 
a minor contribution, in particular for low and middle vibrational levels ($v=0$ 
and $v=10$). (ii) As a consequence of the Franck-Condon overlap, the $^3\Pi$ and 
$^1\Pi$ contributions to the cross section for $v=0$ extends up to 11.5 eV, with 
a maximum around 9 eV. (iii) Beyond 11.5 eV, the asymptotic behavior is driven 
by the $^3\Sigma^-$ resonance. (iv) Sinve high vibrational levels ($v=40$) approach 
the NO dissociation limit, the contributions 
from $^3\Sigma^-$ and, in particular, from the $^1\Sigma^+$ and $^1\Delta$ 
states, become comparable to those from the $^3\Pi$ and $^1\Pi$ resonances at threshold.

Finally, Fig.~\ref{fig:xsecDE} reports the full set of DE cross section results 
resolved over the vibrational ladder. By assuming a Maxwellian distribution for 
the electrons, the corresponding rate coefficient $K_v$ is given, as a function 
of the electron temperature $T_e$, by:
\begin{equation}
K_v(T_e) = \left(\frac{1}{m\,\pi}\right)^{1/2}\,\left(\frac{2}{k_B 
T_e}\right)^{3/2}\,\int\,\epsilon\,\sigma_v(\epsilon)\,e^{-\epsilon/k_B 
T_e}\,d\epsilon\,,
\end{equation}
where $k_B$ is the Maxwell-Boltzmann constant.

In conclusion,   vibrational state-resolved 
cross sections for dissociation of nitric oxide  by electron-impact are computed  for the first time 
using a phenomenological Local-Complex-Potential approach. 
Among the five resonances we considered in the calculations, the $^3\Pi$ 
symmetry is the one which makes the largest contribution.
The full set of data obtained in the present work is available as supplementary 
material to this letter.

\section*{Acknowledgements}

VL and IFS wish to thank the `Institut de recherche Energie Propulsion \& 
Environnement' (I-EPE, CNRS-Normandie Universit\'{e}, France),  la Region Normandie, FEDER, and LabEx-EMC$^3$ \textit{via} the projects EMoPlaF and CO$_2$--VIRIDIS for the financial support. VL thanks the `Lab. Ondes et Milieux Complexes' (CNRS-Universit\'{e} du 
Havre, France), where this work was started, for the kind hospitality.


\end{document}